\documentclass[11pt]{article}
\newlength{\dinwidth}
\newlength{\dinmargin}
\newcommand\eqa{\begin{equation}}
\newcommand\eqb{\end{equation}}
\newcommand\Eqa{\begin{eqnarray}}
\newcommand\Eqb{\end{eqnarray}}
\newcommand{\unit}[1]{\mbox{\it #1}}              
\newcommand{\subs}[1]{\mbox{\scriptsize\it #1}}   

\newcommand\ra{\begin{math}\rightarrow\ \end{math}}  

\newcommand{\eqref}[1]{Eq.~(\ref{#1})}            
\newcommand{\eqRef}[2]{Eqs.~(\ref{#1}, \ref{#2})}
\newcommand{\eqREF}[2]{Eqs.~(\ref{#1}--\ref{#2})}
\newcommand{\fig}[1]{Figure~\ref{#1}}             
\newcommand{\tab}[1]{Table~\ref{#1}}              
\newcommand{\secref}[1]{Section~\ref{#1}}       
 
\newcommand\identity{1\!\!1}
\newcommand\half{\frac{1}{2}}                     
\newcommand\Tr{\mbox{Tr}}                         
\newcommand\gammh[2]{\gamma^{#1}_{#2}}            
\newcommand\gstar[2]{\left.                       
\gammh{#1}{#2}\right.^*}                          
\newcommand\gdagg[2]{\left.                       
\gammh{#1}{#2}\right.^\dagger}
\newcommand\gtran[2]{\left.                       
\gammh{#1}{#2}\right.^T}
\newcommand\gmu{\gamma^\mu}                       
\newcommand\dmu{\partial_\mu}                     

\usepackage{amssymb}
\newcommand\C{\mathbb{C}}                          
\newcommand\Z{\mathbb{Z}}                          
 
\newcommand\Ac{{\cal A}}                          
\newcommand\Bc{{\cal B}}                          
\newcommand\Dc{{\cal D}}                          
\newcommand\Oc{{\cal O}}                          
 
\newcommand\Ab{\bar{A}}                           
\newcommand\Rb{\bar{R}}                           
\newcommand\Phib{\bar{\Phi}}
\newcommand\Gammab{\bar{\Gamma}}
\newcommand\nb{\bar{n}}                           
\newcommand\yb{\bar{y}}                           
\newcommand\mub{{\bar{\mu}}}                      
\newcommand\psib{\bar{\psi}}                      
\newcommand\phib{\bar{\phi}}                      
 
\newcommand\Rt{\tilde{R}}                         
\newcommand\Ut{\tilde{U}}                         
\newcommand\pt{\tilde{p}}                         

\newcommand\muh{\hat{\mu}}                        

\newcommand\eff{_{\protect\subs{eff}}}

\newcommand\gH[1]{\gammh{H}{#1}}
\newcommand\gK[1]{\gammh{K}{#1}}
\newcommand\gstarH[1]{\gstar{H}{#1}}
\newcommand\gstarK[1]{\gstar{K}{#1}}

\newcommand\gdaggK[1]{\gdagg{K}{#1}}

\newcommand\gtranK[1]{\gtran{K}{#1}}
\newcommand\intp{\frac{1}{(2\pi)^d}\int dp \ }
\newcommand\intpc{\int_{\Bc}\frac{dp}{(2\pi/a)^d}}
\newcommand\intpch{\int_{\Bc}\frac{dp'}{(2\pi/a)^d}}
\newcommand\sumk{\sum_{k\in\Z^d}}
\newcommand\kk{\frac{2\pi}{a}k}
\newcommand\HK{H\Delta K}
\newcommand\dlatt{\tilde{\bigtriangleup}}
\newcommand\dellatt{\tilde{\bigtriangledown}}
\begin{document}

\title{
  %
   {\vspace{-4cm} \normalsize \hfill
   \parbox{31mm}{MS-TPI-96-16\\hep-lat/9610029}}\\[20mm]
  %
The perfect action for non-degenerate staggered fermions
} 
\author{%
  H. Dilger%
  \smallskip\\
  Institut f\"ur Theoretische Physik I, WWU M\"unster,\\
  Wilhelm-Klemm-Str.\ 9, D-48149 M\"unster, Germany
}

\maketitle
 
\begin{abstract}
The perfect action of free staggered fermions is calculated by blocking from 
the continuum for degenerate and non-degenerate flavor masses. 
The symmetry structure, connecting flavor transformations and translations, is 
explained directly from the blocking scheme. 
It is convenient to use a modified Fourier transformation, respecting this 
connection, to treat the spin-flavor structure of the blockspins. 
The perfect action remains local in the non-degenerate case; it is explicitly
calculated in two dimensions.
I finally comment on the relation of the blocking scheme to the transition 
from Dirac-K\"ahler fermions to staggered fermions.
\end{abstract}
 

\section{Introduction}
The lattice formulation of fermions is still subject to an intense debate 
under two major questions: The first one concerns the improvement of fermion
actions and operators to reduce lattice artefacts in a systematic way
\cite{improvement}. Secondly the formulation of gauged chiral fermions exhibits
severe problems \cite{Rome}. They also show up in vector-like
theories -- writing down a lattice fermion action, one must choose
between explicit breaking of chiral symmetry, unwanted doublers, or non-local
actions. This is summarized in the no-go theorem of Nielsen and
Ninomiya \cite{Nogo}.

In contrast to lattice gauge fields the formulation of lattice fermions by a 
systematic geometric concept is only given for staggered fermions \cite{BJ}. 
A root of the continuum Laplacian is defined acting on the space of 
inhomogeneous differential forms \cite{Kaehler}. This Dirac-K\"ahler (DK) 
operator is equivalent to the ordinary (free) Dirac operator acting on 
$2^{d/2}$ flavors of spinor fields. The lattice formulation is obtained by the 
transition from forms in the continuum to cochains on the lattice, leading to
the staggered fermion action \cite{Susskind}. It does not produce any doublers,
while preserving part of the chiral symmetry. In view of the above no-go 
theorem this is possible because of the flavor degeneration of the continuum 
theory taken as starting point.

A different way to a better understanding of lattice fermion actions can be
obtained by the renormalization group (RG) \cite{Wilson-RG}. This approach has 
been worked out to a large extent in \cite{W-fermion,W-Wilson}. Starting from a
particular original (nearest neighbor) lattice action, an infinite number of 
RG transformations (RGTs) leads to a perfect action. Provided they preserve the
symmetry structure of the original action, this structure holds in the perfect 
action too. Moreover, if the RGTs show an additional chiral symmetry the 
question of chiral symmetry restoration in the continuum limit can be analyzed 
\cite{W-fermion,W-Wilson}. 
The crucial point is whether the resulting perfect action remains local 
(decreases exponentially). If this is the case, ultra-local actions (with a 
finite range of non-vanishing couplings) might be constructed by truncation of 
the perfect action, with the symmetry structure determined by the corresponding
RGT. 

The RG approach becomes more transparent if the perfect action is directly 
constructed by blocking from the continuum \cite{Wilson-bc}. 
One might directly design a lattice action with a particular 
symmetry structure by an appropriate blocking scheme. The Nielsen-Ninomiya 
no-go theorem decides whether the resulting perfect action can be expected to 
be local. In \cite{W-Wilson} this was studied for the blocking scheme
corresponding to Wilson fermions. For staggered fermions the perfect action
is calculated in \cite{W-fermion} as the fixed point of a lattice blockspin 
transformation, originally proposed in \cite{Kalkreuter}. 

I will calculate the perfect staggered fermion action directly by blocking the 
continuum fermion fields, as proposed already in \cite{Mack}.
After the introduction of the blocking scheme in \secref{scheme} and its 
symmetries in \secref{symmetry} this is performed in \secref{main} for a 
degenerate mass term, corresponding to standard staggered fermions. The main 
technical ingredient is a modified Fourier transformation for the blockspin 
variables, consistent with the discrete remnants of flavor symmetry. 
For two dimensions a perfect action for $2^{d/2}$ fermions was constructed
by blocking from the continuum in a different way, which does not directly 
correspond to staggered fermions \cite{Gabi}.

A generalization of staggered fermions to non-flavor-degenerate mass terms 
is desirable for a physical interpretation of the $2^{d/2}$ flavors.
This was discussed already in \cite{flavors1,flavors2}. In \secref{flavors} I 
calculate the perfect action using the blocking scheme of \secref{scheme}, 
yet with a continuum action with non-degenerate flavor--dependence. 
This perfect action can be shown to be local, and thus may be truncated to 
be used in simulations of flavors with different masses.
In \secref{example} I discuss this action in more detail for two dimensions.
Here I show even-odd decoupling for $m^\dagger m$, $m$ is the fermion matrix.
Therefore numerical calculations using the pseudofermion method 
\cite{Pseudoferm} are possible without an additional flavor doubling, as for 
standard staggered fermions. A detailed study of the four-dimensional case 
shall be subject of a forthcoming paper. 
 
The blocking scheme closely resembles the cochain construction of staggered 
fermions from DK fermions, which can itself be looked at as a blocking
procedure with partial decimation. In fact, the cochain construction served
as a guideline for the staggered fermion blockspin transformation originally
proposed in \cite{Kalkreuter}. In \secref{DK} I discuss this similarity.
However, an attempt to construct an alternative perfect action using this 
cochain construction fails.

\section{The blocking scheme} \label{scheme}

The starting point of the perfect action with staggered fermion symmetry are
$N_f=2^{d/2}$ flavors of continuum fermion fields $\psi_a^b(x)$, with $a$ and 
$b$ the spinor and flavor index, respectively. The action reads
\eqa \label{Scont}
    S[\psib,\psi] \ = \ \int dx \ \psib_a^b(x) \, (\gmu_{aa'} \dmu \ + \ 
                        \delta_{aa'} m) \,\delta_{bb'}\, \psi_{a'}^{b'}(x) \ . 
\eqb
Summation over double indices is understood, $x$ is the $d$-dimensional 
continuous space coordinate. One conveniently defines the component functions
of inhomogeneous differential forms 
\eqa \label{components}
    \Phi \ = \ \sum_H \varphi(x,H) dx^H \ , \ \ 
    dx^H \ = \ dx^{\mu_1} \wedge \dots \wedge dx^{\mu_h} \ , \ \
    \mu_1 < \dots < \mu_h \ .
\eqb
This defines $H$ as a set of $h$ indices (multi-index) $H=\{\mu_1,\dots,
\mu_h\}$. As described in \secref{DK}, this is the starting point for DK 
fermions \cite{Kaehler}, and the whole procedure bears a strong resemblance to 
the mapping of forms onto lattice cochains leading from the continuum DK 
equation to staggered fermions on the lattice \cite{BJ}. I disregard this 
point up to \secref{DK} and treat the following equations as a mere 
prescription for a particular kind of blockspin definition (looking 
unnecessarily complicated).
The $2^{d/2}$ Dirac spinors are unitarily transformed to the component 
functions $\varphi(x,H)$, with $H=\{\mu_1,\dots,\mu_h\}$. Transformation and
backtransformation read
\Eqa
    \varphi(x,H) & = & \frac{1}{\sqrt{N_f}} \sum_{ab} \gstarH{ab} \, 
                       \psi_a^b(x) \ , \ \ \ 
        \gamma^H \ = \ \gamma^{\mu_1} \gamma^{\mu_2} \dots \gamma^{\mu_h}  
                                                            \label{psitophi}\\
    \psi_a^b(x)  & = & \frac{1}{\sqrt{N_f}} \sum_H \gH{ab} \, \varphi(x,H)  \ .
                                                            \label{phitopsi}  
\Eqb
In order to verify the backtransformation use the orthogonality and 
completeness relations 
\Eqa
    \sum_H \gH{ab} \, \gstarH{a'b'} & = & N_f \ \delta_{aa'} \,\delta_{bb'} \ ,
                                                                \label{sumH}\\
    \sum_{ab}\gH{ab}\,\gstarK{ab}\, & = & N_f \ \delta_{HK} \ . \label{sumab}
\Eqb

As further ingredients to the blockspin definition one needs a coarse lattice 
$\Gammab=\{\yb \,|\, \yb_\mu=a\nb_\mu\}$, and a fine lattice $\Gamma = \{y 
\,|\, y_\mu=(a/2)n_\mu\}$, with $\nb_\mu,n_\mu\in\Z$. The fine lattice points 
$y$ are uniquely decomposed as ($e_\mu$ is the vector of length $a$ in 
$\mu$--direction)
\eqa
    y \ = \ \yb \, + \, e_H/2 \ , \ \ e_H \ = \ \sum_{\mu\in H} e_\mu \ .
\eqb 
\newcommand\smallhalf{\mbox{\small /2}}
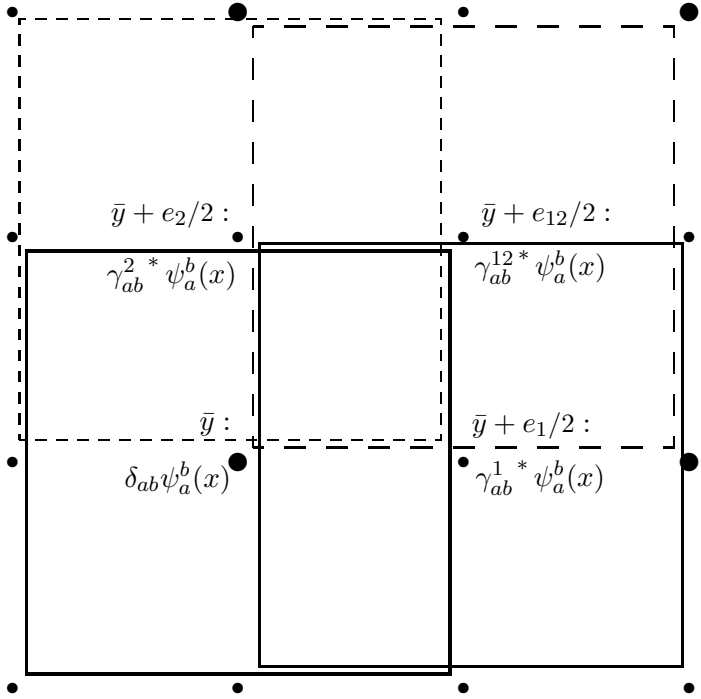
\begin{figure}[t]
\vspace{2mm} 
\unitlength=1mm
\linethickness{0.6pt}
\begin{picture}(100,90)

\put(40,30){\circle*{2.5}}
  {\linethickness{0.9pt}
\put(12,02){\framebox(56,56){}}  }
\put(37,35){\makebox(0,0){$\yb:$}}
\put(32,28){\makebox(0,0){$\delta_{ab}\psi_a^b(x)$}}

\put(70,30){\circle*{1.5}}
\put(43,03){\framebox(56,56){}}
\put(79,35){\makebox(0,0){$\yb+e_1\smallhalf:$}}
\put(80,28){\makebox(0,0){$\gstar{1}{ab}\psi_a^b(x)$}}

\put(40,60){\circle*{1.5}}
\put(11,33){\dashbox{1.4}(56,56){}}
\put(31,63){\makebox(0,0){$\yb+e_2\smallhalf:$}}
\put(31,55){\makebox(0,0){$\gstar{2}{ab}\psi_a^b(x)$}}

\put(70,60){\circle*{1.5}}
\put(42,32){\dashbox{2.8}(56,56){}}
\put(81,63){\makebox(0,0){$\yb+e_{12}\smallhalf:$}}
\put(80,56){\makebox(0,0){$\gstar{12}{ab}\psi_a^b(x)$}}

\put( 40,90){\circle*{2.5}}
\put(100,30){\circle*{2.5}}
\put(100,90){\circle*{2.5}}
\put(100,60){\circle*{1.5}}
\put( 10,00){\circle*{1.5}}
\put( 10,30){\circle*{1.5}}
\put( 10,60){\circle*{1.5}}
\put( 10,90){\circle*{1.5}}
\put( 70,90){\circle*{1.5}}
\put( 40,00){\circle*{1.5}}
\put( 70,00){\circle*{1.5}}
\put(100,00){\circle*{1.5}}
\end{picture}
\vspace{-0mm}
\caption{Blocking scheme for $d=2$. The coarse lattice points $\yb$ are marked 
by larger circles. The blockspin at point $\yb+e_H/2$ is the average of the 
specified projection of $\psi_a^b(x)$ in spin--flavor space, in the block 
surrounding that point.}
\label{block-fig}
\vspace{2mm}
\end{figure}
This defines the multi-index $H(y)$ as position of the fine lattice point $y$
with respect to the coarse lattice $\Gammab$.
The blockspin variables $\phi(y)$ are now defined as averages of the component 
functions $\varphi(x,H(y))$ over the lattice hypercubes $[y]=\{x\ |\ -a/2 \leq 
x_\mu-y_\mu \leq a/2 \}$, as proposed already in \cite{Mack},  
\eqa
    \phi(y) \ = \ \frac{1}{a^d} \int_{[y]} dx \ \varphi(x,H(y)) 
            \ = \ \frac{1}{a^d\sqrt{N_f}} \sum_{ab} \gstar{H(y)}{ab} \, 
                  \int_{[y]} dx \ \psi_a^b(x) \ ,           \label{blockspin}
\eqb
see \fig{block-fig}.
This means for fixed $H$ the component functions are blocked onto the coarse
lattice, the block centers, however, are staggered depending on the multi-index
$H$, leading to one-component blockspins on the fine lattice.
Staggering the block centers is essential. If the coarse grid points were used
as blocking centers for blockspins $\tilde{\phi}(\yb,H)$ this procedure
would commute with the backtransformation \eqref{phitopsi} to the spinor basis,
leading to a separate blocking of the $N_f$ flavors.
As pointed out in \cite{W-fermion} such a blocking scheme leads to a local 
perfect action, only if an additional chiral symmetry breaking term is 
included, whereas with \eqref{blockspin} part of the chiral symmetry survives,
and the perfect action will turn out to be local.

\section{Blockspin symmetry}                                \label{symmetry}

The blocking scheme determines the symmetry of the perfect action simply by
induction of continuum symmetries consistent with the definition of the 
blockspin variables. In particular, from \eqref{blockspin} one is lead to the
symmetries of staggered fermions \cite{flavors2,symmetry}.
It is evident from the block sizes determined by the coarse lattice that the 
blockspin action will be invariant under translations on the coarse lattice.
The restriction of flavor and chiral symmetry is somewhat more involved.
In the continuum these transformations are generated by
\Eqa
    \mbox{flavor transformation: } \qquad \psi'{}_a^b(x) 
    & = & \psi_a^{b'}(x) \gdaggK{b'b}               \label{flavor}\\
%
    \mbox{chiral transformation: } \qquad \psi'{}_a^b(x) 
    & = & \psi_{a'}^b(x) \gamma^5_{aa'} \ .         \label{chiral}
\Eqb
In general flavor transformations are not consistent with the blocking scheme
\eqref{blockspin}. Formally they would induce a transformation of the
blockspins 
\eqa
    \phi'(y) \ = \ \rho(H,K) \ \frac{1}{N_f\,a^d} \sum_{ab} \gstar{H(y) 
                   \Delta K}{ab} \int_{[y]} dx \ \psi_a^b(x) \ . \label{wrong}
\eqb
The sign functions $\rho(H,K)$ are defined such that
\eqa
    \gH{} \, \gK{} \ = \ \rho(H,K) \ \gammh{\HK}{} \ , \label{rho-def}
\eqb
the symmetric difference $\HK=(H\cup K)\backslash(H\cap K)$ fulfills
$H(y \pm y')=H(y)\Delta H(y')$. 
One easily proves
\eqa
    \rho(H\Delta H',K) \ = \ \rho(H,K) \ \rho(H',K) \ , \ \
    \rho(H,K\Delta K') \ = \ \rho(H,K) \ \rho(H,K') \ .     \label{rho-comb}
\eqb
However, $\phi'(y)$ is not a proper blockspin variable, because the multi-index
$H(y) \Delta K$ of the $\gamma$--matrix in \eqref{wrong} is not equal to the
multi-index $H(y)$ of the block center $[y]$. For the discrete transformations 
in \eqref{flavor} (now considered as finite unitary transformations) this can
be cured by a combination with an appropriate fine lattice shift $x$ \ra $x-
e_K/2$. The corresponding transformation of the blockspins becomes 
\eqa
    d^K \, \phi(y) \ = \ \rho(H(y),K) \ \phi(y+e_K/2) \ .   \label{dK-def}
\eqb  
The discrete chiral transformation combined with a shift $x$ \ra $x-e_5/2$
($5\equiv\{1234\}$ for $d=4$) reads 
\eqa
    C \, \phi(y) \ = \ \rho(5,H(y)) \ \phi(y+e_5/2) \ .   \label{C-def}
\eqb
Now consider the combination of chiral transformation \eqref{chiral} and 
discrete flavor transformation given by $\gdagg{5}{bb'}$ in \eqref{flavor}. 
In the $\varphi(x,H)$--basis it simply acts as a multiplication by $(-1)^h$.
Therefore it generates transformations, also defined for the blockspins
\eqa \label{even-odd}
    (c(\alpha)\,\phi)\,(\yb+e_H/2) \ = \ e^{i\alpha(-1)^h} \ 
                                          \phi(\yb+e_H/2)  \ ,
\eqb
corresponding to the continuous even--odd symmetry of staggered fermions.

Since fine lattice shifts must be combined with flavor transformations to yield
a symmetry of the blockspin theory, ordinary Fourier transformation
with momenta chosen in the fine lattice Brillouin zone is not convenient for a 
diagonalization of propagator and action. Instead, the correct basis 
transformation to this end is given by the lattice fields realizing the 
irreducible representations of coarse lattice translations and discrete flavor 
transformations \cite{irreps}. 
They do not commute, thus the irreducible representations (with non-trivial
representation of the sign factor) are multi-dimensional. They are labeled by 
a momentum $p$ in the coarse lattice Brillouin zone $\Bc$: 
$-\pi/a\leq p_\mu<\pi/a$. I call the corresponding basis transformation
symmetry consistent Fourier transformation (scFT)
\eqa
   \phi_a^b(p) \ = \ \sum_y \, e^{ipy} \, \gammh{H(y)}{ab} \, \phi(y) \ .
                                                           \label{ytopab}
\eqb
Using \eqRef{sumH}{sumab} the backtransformation is
\eqa
   \phi(y) \ = \ \frac{1}{N_f} \intpc \ \sum_{ab} \, e^{-ipy} 
   \gstar{H(y)}{ab} \phi_a^b(p) \ .         \label{pabtoy}
\eqb
The discrete (modified) flavor transformations $d^K$ take the form
\eqa
    (d^K\phi)_a^b(p) \ = \ e^{ip\,e_K/2} \ \phi_a^{b'}(p) \ \gdaggK{b'b} \ . 
\eqb
Their meaning, a combined flavor transformation and translation, can be 
read off directly.

For the lattice restriction of rotations and reflections the same 
considerations as for flavor transformations apply. In \eqref{blockspin} the 
transformation of the blocking cell $[y]$ must be accompanied by the 
corresponding transformation of $\gstar{H(y)}{ab}$ by a suitable transformation
in spin and flavor space. This is given by a combination of spinorial rotations
and reflections with flavor transformations, called geometric transformations.
In the $\phi_a^b(p)$ basis they are defined by
\Eqa
    \mbox{geometric rotation by $90^\circ$:}   && R_G^{\mu\nu} \phi_a^b(p) 
    \ = \ \half (1+\gamma^{\mu\nu})_{aa'} \ \phi_{a'}^{b'} (R_{\mu\nu}^{-1}p)
          \ (1-\gammh{\mu\nu}{})_{b'b} \ , \ \qquad       \label{geomRot}\\
    \mbox{geometric reflection:} \,\ \qquad&& \Pi_G^{\mu} \, \phi_a^b(p) 
    \ = \ \gammh{\mub}{aa'} \ \phi_{a'}^{b'}(\Pi^\mu p) \ \gdagg{\mub}{b'b} \ ,
                                                              \label{geomRefl}
\Eqb
with $\mub=\mu\Delta1234$, 
$(R_{\mu\nu}^{-1}p)_\mu=p_\nu,\ (R_{\mu\nu}^{-1}p)_\nu=-p_\mu$, 
and $(\Pi^\mu p)_\nu=(1-2\delta_{\mu\nu})p_\nu$. 
An analogue to spinorial rotations by $180^\circ$ and reflections can be 
defined by combinations with the discrete modified flavor transformations of 
\eqref{dK-def}
\Eqa
    \mbox{spinorial rotation by $180^\circ$:} && \Rb_S^{\mu\nu} \phi_a^b(p) 
    \ = \ \gammh{\mu\nu}{aa'} \ \phi_{a'}^{b} (R_{\mu\nu}^{-2} p) \
          e^{-ip\,e_K/2} \ ,                                  \label{spinRot}\\
    \mbox{spinorial reflection:} \ \ \ \qquad&& \Pi_S^{\mu} \, \phi_a^b(p) 
    \ = \ \gammh{\mub}{aa'} \ \phi_{a'}^{b}(\Pi^\mu p) \ e^{-ip\,e_{\mub}/2}\ .
                                                              \label{spinRefl}
\Eqb
The elimination of the flavor transformation is paid by the combination with
a translation, in order to fit the blocking scheme of \eqref{blockspin}.

\section{The perfect action}  \label{main}

Everything to be done in a free theory is the calculation of the blockspin
propagator $u(y,y')\equiv\langle\phi(y)\phib(y')\rangle$ using the continuum
action $S[\psib,\psi]$, and the definition of the blockspins $\phi(y)$ by the 
continuum fields $\psi_a^b(x)$.
The perfect blockspin action is then given by the inverse propagator $m=u^{-1}$
\eqa
    S\eff[\phib,\phi] \ = \ \sum_{y,y'} \phib(y) \, m(y,y') \, \phi(y') \ ,
                                                               \label{Slatt}
\eqb 
i.e.\ it reproduces $u(y,y')$ in the lattice path integral. Let me first 
evaluate 
\eqa
    \langle \ \frac{1}{a^d} \int_{[y]}  dx  \ \psi_a^b(x) \ 
      \frac{1}{a^d} \int_{[y']} dx' \ \psib_{a'}^{b'}(x') \ \rangle 
    \ = \ \delta_{bb'} \ \Ut_{aa'}(y-y') \ . 
\eqb
It results in
\Eqa
    \Ut_{aa'}(y) 
    & = & \intp e^{-ipy} \ (i\gmu p_\mu\,+\,m)_{aa'} R(m,p) \ ,
                                                            \label{def-Util}\\
    R(m,p)
    & = & \frac{1}{p^2+m^2} \prod_\mu \frac{\sin^2(ap_\mu/2)}{(ap_\mu/2)^2} \ .
                                                            \label{def-R}
\Eqb
Comparing with \eqref{blockspin} the blockspin propagator becomes
\Eqa
    u(y,y')    
%
    & = & \rho(y-y',y') \ U(y-y') \ ,  \qquad           \label{propagator}\\
    U(y) 
    & = & \frac{1}{N_f} \sum_{aa'} \gstar{H(y)}{aa'} 
                          \ \Ut_{aa'}(y) \ .               \label{def-U}
\Eqb
The sign function in \eqref{propagator} $\rho(y-y',y')=\rho(H(y-y'),H(y'))$ is
defined in \eqref{rho-def}. Due to its dependence on $y'$ as second argument it
is not invariant under fine lattice translations but under the discrete flavor
transformations $d^K$ of \eqref{dK-def}. In order to diagonalize propagator 
and action I therefore use the symmetry consistent Fourier transformation 
(scFT) in \eqRef{ytopab}{pabtoy}.

In order to evaluate the perfect action $S\eff$, it is convenient not to invert
the propagator directly. Instead, I calculate $U(y)$ by inversion of (the 
formal expression of) the action, then I compare with \eqRef{def-U}{def-Util}.
Due to its symmetry properties $m(y,y')$ can be written as
\eqa
    m(y,y') \ = \ \rho(y-y',y') \ M(y-y') \ .        \label{My-def}
\eqb
Inserting the inverse scFT \eqref{pabtoy} I obtain
\Eqa
    S\eff 
          & = & \frac{1}{N_f^2} \intpc \intpch \ \sum_{ab} 
                \sum_{a'b'} \ \phib_a^b(p) \, \phi_{a'}^{b'}(p') \nonumber\\ 
     & \times & \sum_z \, e^{ipz} \, \gammh{H(z)}{ac} \, 
                            M(z) \, X_{ca'}^{bb'}(p-p') \ .
\Eqb
With
\eqa
    X_{ca'}^{bb'}(p-p') 
    \ \equiv \ \sum_{y'}\,e^{i(p-p')y'} \,\gammh{H(y')}{cb} \gstar{H(y')}{a'b'}
    \   =    \ N_f \ \delta_{ca'} \ \delta_{bb'} \ (2\pi/a)^d \delta(p-p') \ ,
                                                       \label{orthogonality}
\eqb
following from \eqref{sumH} and the restriction of $p,p'$ to the coarse
Brillouin zone $\Bc$, I arrive at
\Eqa
    S\eff & = & \frac{1}{N_f} \intpc \ \sum_{ab} \sum_{a'b'} \ \phib_a^b(p) \, 
                \phi_{a'}^{b'}(p) \ \delta_{bb'} \, M_{aa'}(p) \ ,
                                                             \label{S-formal}\\
    M_{aa'}(p) & = & \sum_y \, e^{ipy} \, \gammh{H(y)}{aa'} \, M(y) \ .
                                                             \label{Mp-def}
\Eqb  
The propagator of the symmetry consistent Fourier transformed fields is then
\eqa
    \langle \phi_a^b(p) \, \phib_{a'}^{b'}(p') \rangle \ = \ 
    N_f \frac{2\pi^d}{a^d} \ \delta(p-p') \ \delta_{bb'} \ M^{-1}_{aa'}(p) \ ,
\eqb
and with the inverse scFT in \eqref{pabtoy} 
\eqa
    U(y) \ = \ \langle \phi(y) \, \phib(0) \rangle 
         \ = \ \frac{1}{N_f} \intpc \ \sum_{aa'} \, e^{-ipy} \,
               \gstar{H(y)}{aa'} \, M^{-1}_{aa'}(p) \ .
\eqb
One has now the desired equation to compare with \eqRef{def-U}{def-Util}. It 
follows
\Eqa
    M^{-1}_{aa'}(p) & = & \frac{1}{a^d} \, \Bigl( i\gmu_{aa'} Q_\mu(p)
                          \, + \, \delta_{aa'} \, Q_0(p) \Bigr) \ , \\
    Q_\mu(p) & = & \sumk (-1)^{k_\mu} \, (p_\mu+\kk_\mu) \, R(m,p+\kk) \ , \\
    Q_0(p)   & = & m \, \sumk R(m,p+\kk) \ .
\Eqb
Inversion and an inverse scFT of $M_{aa'}(p)$, see \eqref{Mp-def}, lead to
\eqa
    M(y) \ = \ \frac{a^d}{N_f} \intpc \ \sum_{aa'} e^{-ipy} \gstar{H(y)}{aa'} 
               \frac{ -i\gmu_{aa'} Q_\mu(p) \, + \, \delta_{aa'} \, Q_0(p)} 
                    {Q^\mu(p) Q_\mu(p) + Q_0(p)^2} \ .   
\eqb
Using \eqref{sumab} a non-vanishing $M(y)$ appears only for certain positions 
of $y=\yb+e_H/2$ with respect to the coarse lattice points $\yb$.
The final result for the perfect action is therefore
\Eqa
    S\eff[\phib,\phi] 
    & = & \sum_{y,y'} \ \phib(y) \ \rho(y-y',y') \ M(y-y') \ \phi(y') \ , 
                                                          \label{S-result}\\
    M(\yb+e_\mu/2) 
    & = & a^d \intpc\,e^{-ip\yb} \ \ \frac{-iQ_\mu(p)\,e^{-ia\,p_\mu/2}}
          {Q^\mu(p) Q_\mu(p) + Q_0(p)^2} \ ,          \label{S-result-mu}\\
    M(\yb) 
    & = & a^d \intpc \, e^{-ip\yb} \ \ \frac{Q_0(p)}
          {Q^\mu(p) Q_\mu(p) + Q_0(p)^2} \ ,          \label{S-result-0}\\
    M(\yb+e_H/2) 
    & = & 0 \ \mbox{ for } H \neq \mu,\emptyset \ .       \label{S-result-H}
\Eqb
It has the structure of the staggered fermion action \cite{Susskind}. The 
latter arises in the above notation by restriction of $M(y-y')$ to the nearest 
neighbor couplings $M_{KS}(y-y')$
\eqa
    M_{KS}(\pm e_\mu/2) \ = \ \mp \, a^{d-1} \ , \qquad 
    M_{KS}(0)              \ = \   m \, a^d \ .                     \label{KS}
\eqb

The result of \eqREF{S-result}{S-result-H} was achieved in \cite{W-fermion}%
\footnote{
The authors considered a generalization by gaussian smearing of the 
blocking $\delta$-function.
}
as fixed point action of a RGT from a fine to a coarse lattice proposed in 
\cite{Kalkreuter}, see also the third paper of \cite{improvement}.
This RGT commutes with the blocking scheme from the continuum 
to fine and coarse lattice, respectively.
The locality of the perfect action can be read off from the analyticity of the 
Fourier transformed $p$-dependent fractions in \eqRef{S-result-mu}{S-result-0} 
and their periodicity with respect to the coarse Brillouin zone $\Bc$.

\section{Non-degenerate flavors} \label{flavors}

Let me apply the blocking scheme of \secref{scheme} to the case of 
non-degenerate flavors. I assume a continuum propagator with a flavor-dependent
mass
\eqa \label{non-deg}
    \langle \psi_a^b(x) \, \psib_{a'}^{b'}(0) \rangle \ = \ \delta_{bb'} \ 
    \intp \, e^{-ipx} \, \frac{(i\gmu p_\mu\,+\,m_b)_{aa'} }{p^2\,+\,m_b^2} \ .
\eqb
The modification of staggered fermions to the latter case is discussed in 
\cite{flavors1,flavors2}. The perfect action derived here might be used to
explain the necessary structure of such a modification.
I proceed in generalization of \secref{main} with the definition
\eqa
    \langle \ \frac{1}{a^d} \int_{[y]}  dx  \ \psi_a^b(x) \ 
      \frac{1}{a^d} \int_{[y']} dx' \ \psib_{a'}^{b'}(x') \ \rangle 
    \ = \ \sum_K \gK{bb'} \ \Ut^K_{aa'}(y-y') \ , 
\eqb
\eqa
    \Ut^K_{aa'}(y) \ = \ \intp e^{-ipy} \ \frac{1}{N_f} \sum_b \gstarK{bb}  
    (i\gmu p_\mu\,+\,m_b)_{aa'} \, R(m_b,p) \ .              \label{def-UtilK}
\eqb
Since the propagator is diagonal in flavor space $\Ut^K_{aa'}(y)$ is non-zero 
iff $\gK{}$ is diagonal, denoted by $K\in\Dc$. In the following this 
restriction is understood for the sums over $K$. Corresponding to 
\eqRef{propagator}{def-U} I find
\Eqa
    u(y,y')
    & = & \sum_K \rho^K(y') \rho(y-y',y') \ U^K(y-y') \ , \label{propagatorK}\\
    U^K(y)    
    & = & \frac{1}{N_f}\sum_{aa'} \gstar{H(y)}{aa'} \ 
          \Ut_{ac}^K(y)\gtranK{ca'} \ , \\ 
    \rho^K(y') 
    & = & \rho(K,H(y')) \, \rho(H(y'),K) \ .      
\Eqb
Inserting \eqref{def-UtilK} $U^K(y)$ is cast in the form of an inverse 
scFT 
\Eqa
    U^K(y)
    & = & \frac{1}{N_f} \intpc\sum_{aa'} \gstar{H(y)}{aa'} \, e^{-ipy} \ 
          U^K_{aa'}(p) \ ,                                   \label{def-UK}\\
    U^K_{aa'}(p) 
    & = & \frac{1}{a^d} \left( \sum_\mu i(\gmu\gtranK{})_{aa'} \, Q^K_\mu(p)
          \ + \ \gtranK{aa'} Q^K_0(p) \right) \ ,             \label{def-UpK}
\Eqb 
with 
\Eqa
    Q^K_\mu(p) & = & \sumk \ \prod_{\nu\in K\Delta\mu} \!\! (-1)^{k_\nu} \, 
                     \frac{1}{N_f} \sum_b\gstarK{bb} \,(p_\mu\!+\!\kk_\mu)
                     \, R(m_b,p\!+\!\kk) \ ,                 \label{QmuK}\\ 
    Q^K_0(p)   & = & \sumk \ \ \prod_{\nu\in K} (-1)^{k_\nu}
                     \frac{1}{N_f}\sum_b \gstarK{bb} \, m_b
                     \, R(m_b,p\!+\!\kk) \ .               \label{Q0K}
\Eqb
The propagator of the symmetry consistent Fourier transformed fields becomes
\Eqa
    \langle \phi_a^b(p) \, \phib_{a'}^{b'}(p') \rangle
    & = & \sum_K \ \sum_z e^{ipz} \gammh{H(z)}{ac} U^K(z) \ 
          \sum_{y'} e^{i(p-p')y'} \gammh{H(y')}{cb} \gstar{H(y')}{dd'} \ 
          \gdaggK{a'd} \gK{d'b'} \nonumber\\
    & = & N_f (2\pi/a)^d \, \delta(p-p') \ \sum_K \gK{bb'} \ U^K_{ac}(p)
                                                           \, \gstarK{ca'} \ ,
\Eqb
with use of \eqref{orthogonality} and the backtransformation of \eqref{def-UK}
for the last line. Finally I obtain
\Eqa
    \langle \phi_a^b(p) \, \phib_{a'}^{b'}(p') \rangle
    & = & N_f (2\pi/a)^d \, \delta(p-p') \ \sum_K \gK{bb'} \ V^K_{aa'}(p) \ ,
                                                           \label{u-diag}\\
    V^K_{aa'}(p) 
    & = & \frac{1}{a^d} \left( \sum_\mu i\gmu_{aa'} \, Q^K_\mu(p)
          \ + \ \delta_{aa'} Q^K_0(p) \right) \ .             \label{def-VpK}
\Eqb 
Thus, starting with diagonal $\gamma^K$--matrices in flavor space, $K\in\Dc$, 
the diagonality is recovered for $\langle \phi_a^b(p) \phib_{a'}^{b'}(p')
\rangle$.

The action may be written due to coarse lattice translation symmetry 
\Eqa
    S\eff[\phib,\phi] & = & \sum_{y,y'} \phib(y) \, m(y,y') \, \phi(y') \ , 
                                                              \label{SlattK1}\\
    m(y,y') & = & \sum_K \rho^K(y')\rho(y-y',y') \ M^K(y-y')\ . \label{SlattK2}
\Eqb
The transition to the symmetry consistent Fourier transformed fields yields
\Eqa
    S\eff 
    & = &      \frac{1}{N_f} \intpc \ \sum_{ab}\sum_{a'b'} \ \phib_a^b(p) \,  
               \phi_{a'}^{b'}(p) \ \sum_K \, \gK{bb'} \ 
               M^K_{ac}(p) \gstarK{ca'} \ ,                 \label{S-formalK}\\
    M^K_{ac}(p) 
          & = & \sum_z \, e^{ipz} \, \gammh{H(z)}{ac} \, M^K(z) \ .
                                                             \label{MpK-def}
\Eqb
As propagator and action are diagonal in flavor space one has to sum over
diagonal $\gamma^K$--matrices only. This corresponds to the remaining discrete 
flavor symmetry transformations $d^K, K\in\Dc$, see \cite{flavors1}.
It follows that choosing $\Dc=\{\emptyset,12,34,1234\}$ also the geometric 
rotations $\omega^{12},\omega^{34}$ of \eqref{geomRot} remain as a symmetry, 
whereas in general the rotation and reflection symmetry is kept only in the 
modified spinorial form of \eqRef{spinRot}{spinRefl}.    
Inversion of $\sum_K\gK{bb'}M^K_{ac}(p)\gstarK{ca'}$ and comparison with 
\eqref{u-diag} lead to
\eqa 
    \sum_K \rho(K,L) \left( M^K(p) \gtranK{} V^{K\Delta L}(p) \right)_{aa'} 
    \ = \ \delta_{L,\emptyset} \ \delta_{aa'}\ ,\ \ K,L\in\Dc \ .\label{SK-res}
\eqb
This determines $M^K(p)$ and $S\eff$ with use of the backtransformation of 
\eqref{MpK-def}
\eqa
    M^K(\yb+e_H/2) \ = \ \frac{1}{N_f} \intpc \, e^{-ip\yb} \ 
    \Tr \left( e^{-ipe_H/2} \gstarH{} M^K(p) \right) \ . \label{SK-res2}
\eqb
Note that the blockspin propagtor $u(y,y')$ in the non-flavor-degenerate case 
\eqref{propagatorK}, and thus the fermion matrix $m(y,y')$, get complex values 
depending on the multi-indices $H(y),H(y')$. 

Let me consider the vicinity of $p=0$. The factor $R(m,p\!+\!\kk)$ becomes 
(up to corrections of $\Oc(p^2)$) $(\prod_\mu\delta_{k_\mu,0})/(p^2+m^2)$, 
and the sums $\sumk$ in $V^K(p)$ drop out. In this approximation \eqref{sumH} 
can be used to simplify \eqRef{u-diag}{def-VpK}
\eqa
    \langle \phi_a^b(p) \, \phib_{a'}^{b'}(p') \rangle \ = \ N_f (2\pi/a)^d \, 
    \delta(p-p') \ \frac{1}{a^d} 
    \frac{ (\sum_\mu i\gmu\,p_\mu + m_b)_{aa'} }{ p^2+m_b^2 } \delta_{bb'} \ ,
\eqb 
and the Fourier representation of $S\eff$ in \eqref{S-formalK} becomes 
\eqa
    S\eff \ = \ \frac{1}{N_f} \intpc \ \sum_{ab}\sum_{a'b'} \ \phib_a^b(p) \,  
                \phi_{a'}^{b'}(p) \ a^d \Bigl[ \delta_{bb'} 
                (-i\gamma^\mu p_\mu + m_b)_{aa'} + \Oc(p^2) \Bigr] \ .
\eqb

In order to prove $S\eff$ to be local, let me define the transformations 
$B^\mu$ in momentum and spinor space ($\muh$ is the unit vector in 
$\mu$--direction of momentum space)
\eqa
    B^\mu F_{ab}(p) \ = \ \gdagg{\mub}{aa'} \, 
    F_{a'b'}\,(p\!+\!\frac{2\pi}{a}\muh) \ \gammh{\mub}{b'b} \ ,
\eqb
$ \mub = 1234\Delta\mu \ (12\Delta\mu) $ for $ d=4 \, (2) $. 
Since $(B^\mu)^2 V^K(p)=V^K(p)$ thus $(B^\mu)^2 M^K(p)=M^K(p)$, one may 
decompose $M^K(p)$ with respect to its behavior under these transformations 
\eqa
    M^K(p) \ = \ \sum_H m^K_H(p) \ , \qquad \mbox{ with } \ 
    B^\mu \, m^K_H(p) \ = \ \sigma^\mu_H \, m^K_H(p) \ .    
\eqb
Here $\sigma^\mu_H=-1\,(1)$ for $\mu\in H\,(\mu\not\in H)$, and the components 
$m^K_H(p)$ are uniquely determined. With 
$B^\mu\,\gamma^K=\sigma^\mu_K\,\gamma^K$ and 
$B^\mu\,V^K(p)=\sigma^\mu_K\,V^K(p)$ 
it follows from \eqref{SK-res}
\eqa
    \sum_K\rho(K,L)\sum_H\sigma^\mu_{H}\,\sigma^\mu_{L} \left(m^K_H(p)\gtranK{}
    V^{K\Delta L}(p) \right)_{aa'} \ = \ \delta_{L,\emptyset} \ \delta{aa'} \ .
\eqb
This is solved by $m^K_H(p)=\delta_{H,\emptyset}\,M^K(p)$ giving back 
\eqref{SK-res}. Thus $M^K(p)$ is invariant under $B^\mu$, as is the term 
$e^{-ipe_H/2}\gstarH{}$ in the trace of \eqref{SK-res2}, and this trace is 
periodic with respect to the Brillouin zone $\Bc$. On the other hand $M^K(p)$ 
is analytic for all $p\in\Bc$, because $V^K(p)$ has no zeros in $\Bc$. 
In conclusion the perfect action remains local in the non-degenerate case%
\footnote{
After completion of this paper a calculation of the perfect action of 
degenerate staggered fermions by blocking from the continuum was published in 
[16]. 
The authors used a smeared blockspin transformation within a somewhat different
calculation scheme. Optimizing the additional smearing parameters the 
exponential decay constant of the couplings can be increased considerably. 
This can be performed also in the non-degenerate case. For the same choice of 
smearing parameters the impact on the locality is roughly the same as in [16].
}.
I will analyze the structure of this action in a simple case in the following 
section.
 
\section{The two-dimensional case} 
\label{example}

As example I study the case of $d=2$ with fermion masses $m_1,m_2$.  I choose 
$\gamma^{12}=\unit{diag}(i,-i)$, thus $\Dc=\{\emptyset,\,12\}$, and I obtain
\Eqa
    V^\emptyset_{aa'}(p) 
    & = & \frac{1}{2a^d} \sumk  \Bigl[ \Rt^+(\pt) \, \delta_{aa'} \ + \
          \Rb^+(\pt) \sum_\mu (-1)^{k_\mu}\,i\gmu_{aa'}\,\pt_\mu \Bigr] \ , \\
    V^{12}_{aa'}(p) 
    & = & \frac{-i}{2a^d} \sumk \Bigl[ \Rt^-(\pt) \, \delta_{aa'} \ + \
          \Rb^-(\pt) \sum_\mu (-1)^{k_\mu}\,i\gmu_{aa'}\,\pt_\mu \Bigr] 
          \, \epsilon(k) \ ,
\Eqb 
with $\pt=p+\kk$, $\Rb^\pm(\pt)=R(m_1,\pt)\pm R(m_2,\pt)$, 
$\Rt^\pm(\pt)=m_1 R(m_1,\pt)\pm m_2 R(m_2,\pt)$, and 
$\epsilon(k)=(-1)^{k_1+k_2}$. For simplicity of notation I furthermore define
\Eqa 
    V^\emptyset(p) & = & A \ = \ \gamma^\mu a_\mu + \identity a_0 \ , \ \
    V^{12}(p)      \ = \ B \ = \ \gamma^\mu b_\mu + \identity b_0 \ , \\
    (a,b)          & = & a_1 b_1 + a_2 b_2 - a_0 b_0 \ , \ \
    (a,\epsilon b) \ = \ a_1 b_2 - a_2 b_1 \ , \\
    c_\mu          & = & a_\mu b_0 - a_0 b_\mu \ .
\Eqb
With $\Ab=\gamma^\mu a_\mu - a_0$ it follows $\Ab A = (a,a) \equiv a^2$, and
$ \Ab B \ = \ (a,b)\identity + (a,\epsilon b)\gamma^{12} + c_\mu\gamma^\mu $.
Now \eqref{SK-res} reads
\eqa
    M^\emptyset(p)\, A \ + \ M^{12}(p)\gtran{12}{}\, B \ = \ \identity\ ,\qquad
    M^\emptyset(p)\, B \ - \ M^{12}(p)\gtran{12}{}\, A \ = \ 0 \ .
\eqb
This is solved by
\Eqa
    M^\emptyset(p) & = & N^{-1} \Ab \ , \\
    N & = & \Ab A + \Ab B A^{-1} B \ = \ a^2 + \frac{1}{a^2}(\Ab B)^2 \ , \nonumber\\
        & = & \left[ a^2 + \frac{(a,b)^2 - (a,\epsilon b)^2 + c^\mu c_\mu}
              {a^2} \right] \identity 
        \ + \ \frac{2(a,b) (a,\epsilon b)}{a^2} \, \gamma^{12} 
        \ + \ \frac{2(a,b) c_\mu}{a^2} \, \gamma^\mu \nonumber\\
    &\equiv & n_0\,\identity + n_{12}\,\gamma^{12} + n_\mu \gamma^\mu \ , \\
    N^{-1} 
        & = & \frac{1}{Z} ( n_0\,\identity - n_{12}\,\gamma^{12} 
                                             - n_\mu \gamma^\mu ) \ , \qquad
    Z   \ = \ n_0^2 + n_{12}^2 - n_\mu n^\mu  \ ,
\Eqb
for $M^{12}(p)\gtran{12}{}$ interchange $A$ and $B$. Altogether I find
(noting $a_\mu \epsilon^\mu_\nu n^\nu + n_{12} a_0 = 0$)
\Eqa
    M^\emptyset(p) 
    & = & \frac{1}{Z} \Bigl[ (n_0 a_\mu + a_0 n_\mu) \gamma^\mu
    \ + \ n_{12} a_\mu \epsilon^\mu_\nu \gamma^\nu 
    \ - \ (n_\mu a^\mu + n_0 a_0) \identity \Bigr] \ ,     \label{M0-res}\\
    M^{12}(p)
    & = & -\left. M^\emptyset(p) \right|_{a\leftrightarrow b} \ \gamma^{12} \ .
                                                           \label{M12-res}
\Eqb

In \tab{results-tab} the fermion matrix $m(y,y')$ following from 
Eqs.~(\ref{SlattK2}),(\ref{SK-res2}),(\ref{M0-res}),(\ref{M12-res}) is 
evaluated for $0\leq y_\mu<3,y'=0$ and vice versa, putting $m_1=0.2$, $m_2=1$.
The contributions of $M^\emptyset(y-y')$ are real, the contributions of 
$M^{12}(y-y')$ are purely imaginary. The vanishing of the $M^{12}$--part for 
couplings $y-y'= (n/2)e_\mu$, i.e.\ $\Pi^\mu (y-y') = R_{\mu\nu}^{-2}(y-y')$ 
follows from the behavior under geometric rotations $(R^{\mu\nu}_G)^2$ by 
$180^\circ$, which are symmetry transformations, and geometric reflections 
$\Pi^\mu_G$, see \secref{symmetry}. 
Under $\Pi^\mu_G$ the $M^{12}$-contributions pick up a minus sign, because 
this is the part of the action violating the $d^1,d^2$ flavor transformations 
and thus the geometric reflections. 

\begin{table}
\begin{center}
\begin{tabular}{|c|l l l l|}
\hline
 $m(y,0),m(0,y)$ & $y_1=0\qquad$ & $y_1=1/2$ & $y_1=1$ & $y_1=3/2$ \\
\hline
 $y_2=0$   &$\ \ 0.814$ &$ \mp1.915$        &$ -0.054$    &$\pm0.308$ \\
 $y_2=1/2$ &$\mp1.915$ &$\mp 0.172\,i$ &$\pm0.118+0.004\,i$ &$\pm0.048\,i$ \\ 
 $y_2=1$   &$ -0.054$ &$\pm0.118-0.004\,i$ &$\ \ 0.002$&$\mp0.005+0.001\,i$ \\ 
 $y_2=3/2$ &$\pm0.308$ &$\pm0.048\,i$  &$\mp0.005-0.001\,i$ &$\mp0.007\,i$ \\ 
 $y_2=2$   &$ -0.003$ &$\pm0.006+0.003\,i$ &$ -0.001$&$\mp0.002-0.0004\,i$ \\ 
 $y_2=5/2$ &$\mp0.046$ &$\mp0.011\,i$  &$\mp0.003+0.0002\,i$ &$\pm0.0002\,i$ \\ 
\hline
\end{tabular}
\end{center}
\vspace{-2mm}
\caption{The first couplings of the fermion matrix $m(y,0),m(0,y)$.
If forward and backward coupling differ, the upper sign belongs to $m(y,0)$. 
\label{results-tab}}
\vspace{3mm}
\end{table}

It can be read off from \tab{results-tab} that the fermion matrix decomposes
into a hermitian part $m_+$ coupling even sites with even sites, odd sites with
odd sites, and an anti-hermitian part $m_-$ coupling even sites with odd sites.
The sign structure in \eqref{SlattK2} leads to 
\eqa
    m_+(y,y-z)  \ m_-(y-z, y') \ = \ m_-(y,y'+z) \ m_+(y'+z,y') \ , 
\eqb
where it is crucial that $M^\emptyset(\yb+e_{12}/2) = M^{12}(\yb) = 0$. It 
follows $m_+m_-=m_-m_+$ and
\eqa
   m^\dagger\,m \ = \ m_+\,m_+ \ - \ m_-\,m_- \ ,
\eqb
i.e.\ $m^\dagger m$ does not couple even and odd sites. This property is 
useful in numerical simulations with help of the pseudofermion method 
\cite{Pseudoferm}.

\section{Relation to the cochain construction of staggered fermions} \label{DK}
Here I will very shortly present the main features of Dirac-K\"ahler (DK) 
fermions \cite{Kaehler} and their triangularization as lattice cochains. 
For a detailed description I refer to \cite{BJ}. The DK equation 
\eqa
    (\,d \ - \ \delta \ + \ m\,) \ \Phi \ = \ 0 
\eqb
is equivalent to the Dirac equation for $N_f=2^{d/2}$ degenerated flavors.  
$\Phi=\sum_H\varphi(x,H)dx^H$ is a inhomogeneous differential form, see
\eqref{components}, $d$ is the external differentiation, $\delta$ the 
codifferentiation operator. The corresponding action is
\eqa \label{DKaction}
    S_{DK}[\Phib,\Phi] \ = \ [\Phib, (d-\delta+m)\Phi] \ , \qquad \mbox{ with }
    [\Phi,\Phi'] \equiv \sum_H \varphi(x,H) \, \varphi'(x,H) \ .
\eqb
The unitary transformation in \eqref{phitopsi} of the component functions
$\varphi(x,H)$ to the Dirac basis $\psi_y^b(x)$ leads back to the standard 
action \eqref{Scont}.

A form may be considered as a mapping of all $h$--dimensional ($h=1,\dots,d$) 
areas $\Ac$ into $\C$
\eqa
    \Phi(\Ac) \ = \ \int_{\Ac} \Phi \ . 
\eqb
The lattice restriction of these forms arises by restriction of the integration
areas $\Ac$ to lattice chains, i.e.\ combinations of $h$--dimensional lattice 
cells $[\yb,H]$ (sites, links, plaquettes, \dots). The cell $[\yb,H]$ is 
spanned by the coarse lattice unit vectors $e_\mu, \mu\in H$ at the point 
$\yb\in\Gammab$.
It is natural to represent this cell by the fine lattice point at its center 
$y=\yb+e_H/2$. This decomposes the space of all forms into cochains, i.e.\
classes characterized by
\eqa
    \int_{[\yb,H]} \Phi \ = \ \phi(\yb+e_H/2) \ .
\eqb 
Since the boundary of a lattice cell is again a lattice cell, the decomposition
is consistent with the external differentiation $d$. From Stokes' theorem one 
obtains ($\Delta[\yb,H]$ is the oriented boundary of $[\yb,H]$)
\eqa \label{dlatt-def}
    \int_{[\yb,H]} d\Phi \ = \ \int_{\Delta[\yb,H]} \Phi
                         \ \equiv \ (\dlatt\phi)\,(\yb+e_H/2) \ ,
\eqb
i.e.\ if $\Phi,\Phi'$ are of the same class, so are $d\Phi,d\Phi'$.
For the codifferentiation operator such a construction is not that clear, see
\cite{BJ}. However, its lattice correspondence $\dellatt$ may be simply defined
as the adjungated operator with respect to the lattice scalar product $(,)$, in
order to preserve the usual anti-hermiticity property of $\dlatt - \dellatt$.

The transition from continuum forms to lattice cochains can be written by a 
blocking operator $C$ mapping the component functions $\varphi(x,H)$ onto the 
blockspin variables $\phi(y)=\phi(\yb+e_H/2)$
\Eqa
    (C\varphi)(y) 
    & = & \int_{[\yb,H]} \varphi(x,H) 
    \ = \ \int_{[y]} \chi_y(x,H) \, \varphi(x,H) \ , \label{DKblock1}\\
    \chi_y(x,H)
    & = & \prod_{\mu\in H} (1/a) \ \prod_{\mu\not\in H} \delta(x_\mu-y_\mu) \ .
                                                  \label{DKblock2}  
\Eqb
This representation shows the similarity to the blocking scheme in 
\eqref{blockspin}, $[y]$ denotes the full coarse lattice hypercube with center 
$y$. The difference, however, is that the component functions $\varphi(x,H)$ 
are now averaged only over those directions $\mu$ with $\mu\in H$, in the 
other directions the blocking scheme corresponds to decimation 
\cite{decimation}, as indicated by the characteristic functions $\chi_y(x,H)$. 
Now \eqref{dlatt-def} reads $C\,d=\dlatt\,C$, and the lattice restriction of 
the DK action \eqref{DKaction} might be written
\Eqa
    S_{DK}[\Phib,\Phi] & = & [\Phib,d\Phi] - [d\Phib,\Phi] + m[\Phib,\Phi] \\
%
    \longrightarrow  \qquad S[C\Phib,C\Phi] 
    & = & (C\Phib,Cd\Phi) - (Cd\Phib,C\Phi) + m(C\Phib,C\Phi) \nonumber\\ 
    & = & (C\Phib,(\dlatt - \dellatt + m)C\Phi) 
\Eqb
As described in \cite{BJ}, the result $S[\phib,\phi]$ is the staggered fermion 
action of \eqref{KS}.  

One may ask whether the cochain blocking given by $C$ in \eqref{DKblock1}
is a reasonable alternative blockspin definition in the sense of blocking from 
the continuum.
Unfortunately, as prescription in this scheme, cochain blocking violates the
discrete flavor transformation symmetry of staggered fermions, which has
proven quite useful for the calculation of the perfect action. Consider for 
instance the propagator of two blockspins with the same spin-flavor content
\eqa
    u_H(\yb) \ = \ \langle \phi(\yb+e_H/2) \, \phib(e_H/2) \rangle \ .
\eqb
For its evaluation along the lines of \eqref{main} one has to integrate the 
continuum fields over different lattice cells of dimension $h$ depending on 
$H$. This destroys the discrete modified flavor symmetry, which in this case 
requires $H$--independence of $u_H(\yb)$. So, in this naive way, a direct 
exploitation of the cochain blocking scheme seems not to be useful for RG 
considerations.
Nevertheless it seems me worthwhile to look for a more clever combination of
these two approaches to a lattice fermion formulation.

\subsection*{Acknowledgement}
Thanks to G. Schierholz, D. Talkenberger, C. Wieczerkowski, and U.-J. Wiese 
for encouraging discussions. I'm deeply indebted to G. Mack for explaining to 
me the idea of lattice fermions blocked from the continuum, and to H. Joos for 
introducing me into the DK fermion formulation. The financial support by 
the Deutsche Forschungsgemeinschaft is gratefully acknowledged.


\end{document}